# A generalized nonlinear Schrödinger equation as model for turbulence, collapse, and inverse cascade

Dian Zhao (赵典) and M. Y. Yu (郁明阳)

*Institute for Fusion Theory and Simulation, Department of Physics, Zhejiang University,*

*Hangzhou 310027, China*

E-mail:dzhaobx@gmail.com

**Abstract**

A two-dimensional generalized cubic nonlinear Schrödinger equation with complex coefficients for the group dispersion and nonlinear terms is used to investigate the evolution of a finite-amplitude localized initial perturbation. It is found that modulation of the latter can lead to side-band formation, wave condensation, collapse, turbulence, and inverse energy cascade, although not all together and nor in that order.

## I. Introduction

Since many decades the phenomenon of collapse has been encountered in many theoretical studies of nonlinear wave interaction [1–5]. When there is strong modulation or interaction of finite-amplitude waves, there can appear within a finite time a catastrophic decrease of the spatial size of the wave structure, together with extreme concentration of the wave energy. It can occur for gravity waves on surfaces of deep water [6], Langmuir waves in plasmas [1,2,7], and during intense light propagation in nonlinear media [8]. Collapse-like behavior can also appear in other systems such as Bose-Einstein condensates and matter waves [9,10], as well as during gravitational compression of massive stars [11].

Many studies of collapse are based on the nonlinear Schrödinger equation (NLSE), which has become a paradigm model for investigating weakly-nonlinear dispersive waves [1–5]. For most applications, the system of interest consists of the classical Schrödinger equation, but with a nonlinear term, which is self-consistently determined by the nonlinear response of the system. The collapse behavior appears as a mathematical singularity during the evolution of the system. Just before reaching the singularity, or collapse, the wave energy density approaches infinity and the space scale of the wave or structure approaches zero. Collapse thus represents a limitation of the original physical model. One usually invokes a new process, such as nonlinear damping, that might prevent collapse [1–5]. However, it has been shown that many common dissipative and disordering effects, such as linear damping, wave emission, thermal fluctuations, incoherence, non-locality, etc. [1–5, 12–16] cannot fully prevent collapse. In this case it can be expected that the latter will result in a completely new physical state of the system [1–4, 9–11].

There have been many investigations of the standard NLSE and its variants. Most of the investigations are on the formation and properties of solitons and other localized structures, as well as their interactions. Pereira and Stenflo [17] considered a cubic NLSE equation with complex coefficients. The equation has since been extended to include other types of nonlinearities and higher dimensions and exotic soliton, vortex, pattern, and other solutions have been found and investigated [18–21]. Here, we shall investigate the wave collapse behavior using as simple model a two-dimensional generalized cubic NLSE equation. The latter is the standard cubic NLSE, but with complex coefficients for the group-dispersion and nonlinear terms. That is, viscous effects and nonlinear damping/growth, as well as reaction-diffusion phenomena, can be included. We are interested in the nonlinear evolution of a spatially localized finite-amplitude perturbation, in particular the onset of collapse and/or turbulence, as well as what happens after the collapse. Numerical results show that when collapse does not occur, the initial perturbation first undergoes modulational instability, followed by gradual cascading of the wave energy towards that of the shorter waves until a turbulent state is formed. However, when collapse occurs during the modulation, the energy is abruptly transferred to the very short waves, followed by inverse cascade of energy from the resulting short waves to the longer and longer waves, and eventually a spiky turbulent state appears. However, for the cases studied no self-organization to steady state is found.

## II. The generalized NLSE

The generalized cubic NLSE in our model is

$$i\partial_t E + p\nabla^2 E + \left(V(x,y) - q|E|^2\right)E = 0, \qquad (1)$$

where $E(x, y, t)$ is a complex function of the time $t$ and space $(x, y)$, and $\nabla^2 = \partial_x^2 + \partial_y^2$. Unlike the standard cubic NSE, here the coefficients $p\ (= p_r + ip_i)$ and $q\ (= q_r + iq_i)$ can be complex, and an externally given potential $V(x,y)$ has been included. In many physical applications, such as wave interaction in continuous media, Eq. (1) describes the nonlinearly modulated envelope of a wave train. In this case, $V(x,y)$ characterizes an externally given potential, $p_r$ and $q_r|E|^2$ are the coefficients of group dispersion and nonlinear frequency shift, and $p_i$ and $q_i|E|^2$ are the coefficients of viscous damping or growth and nonlinear damping or growth, respectively. We note that $p_i$ and $q_i$ represent wavelength- and amplitude-dependent damping or growth, respectively. That is, the linear Schrödinger operator now has dispersive, diffusive, as well as dissipative parts. As a result of the complex coefficients, the generalized NLSE can describe other physical phenomena, and should be especially useful for investigating the wave cascade process under realistic conditions.

We first look at the conservation properties of the cubic NLSE in the presence of complex coefficients. Multiplying (1) by the complex conjugate $E^*$, we obtain

$$iE^*\partial_t E + pE^*\nabla_\perp^2 E + V|E|^2 - q|E|^4 = 0. \qquad (2)$$

Equation (2) subtracting its complex conjugate results in

$$\partial_t \int_{-\infty}^{+\infty} |E|^2 dV = 2\int_{-\infty}^{+\infty} \left(p_i + q_i|E|^2\right)|E|^2 dV, \qquad (3)$$

where the integration is over all space. Thus the quantity $\int_{-\infty}^{+\infty} |E|^2 dV$, which following convention shall be referred to as the total energy, is conserved if $p_i = 0$ and $q_i = 0$. This is of course expected since Eq. (1) is then the standard cubic NSE [1]. If $|E|^2 \ll 1$, Eq. (3) can be simplified to $\partial_t \int_{-\infty}^{+\infty} |E|^2 dV \sim 2p_i \int_{-\infty}^{+\infty} |E|^2 dV$, so that the total energy grows or damps according $p_i >$ or $< 0$. It is of interest to note that in the former case, the growth of small amplitude modulation will saturate when $|E|^2 = p_i/|q_i|$ if $q_i < 0$, because of the built-in amplitude-dependent damping.

Similarly, multiplying (1) by $\nabla E^* \cdot \nabla$ and subtracting the complex conjugate of the resulting equation, one finds for the evolution of the reduced total enstropy

$$\partial_t \int_{-\infty}^{+\infty} |\nabla E|^2 dV = 2\int_{-\infty}^{+\infty} \left( p_i |\nabla^2 E|^2 + q_i |E\nabla E|^2 \right) dV, \quad (4)$$

so that as expected the reduced total enstropy is also conserved if $p_i = 0$ and $q_i = 0$. Unlike the total energy, in the $|E|^2 \ll 1$ limit the evolution of the total enstropy depend on both the divergence and magnitude of $\nabla E$, as well as the signs of $p_i$ and $q_i$ and the magnitude of $E$. For example, in the $p_i > 0$, $q_i < 0$ case mentioned above, the saturation (if any) level of the total enstropy would be $\left| E\nabla E / \nabla^2 E \right|^2 = p_i / |q_i|$.

### III. Numerical Results

We are interested in the effect of a localized finite-amplitude initial perturbation, the evolution of the resulting modulational instability, as well as the onset of turbulence, wave collapse, and inverse cascade.

We shall solve Eq. (1) numerically using the spectral method [1–5]. The size of the computation grid is $256 \times 256$, and the length of the simulation box on each side is $4\pi$ ($-2\pi$ to $2\pi$). Periodic boundary conditions are used on both sides. The grids are sufficiently fine to ensure that the use of the spectral method does not lead to preferential alignment of the wave and turbulence structures. For the cases considered, aliasing effects are found to remain at a negligible level. There is also good conservation of the total energy and enstropy for the non-dissipative case $(p_i, q_i) = 0$.

For convenience, we shall adopt the external potential

$$V(x, y) = V_0 \cos\left[ \left(\frac{x}{4}\right)^2 + \left(\frac{y}{4}\right)^2 \right], \quad (6)$$

where $V_0 = -10$. The 2D profile of the potential is shown in Fig. 1. As expected, since our system is not conservative and we are interested in finite-amplitude perturbations, the form and magnitude of the potential does not significantly affect the main conclusions. Instead, the size of the simulation box with respect to that of the initial pulse does play a role in triggering the onset of turbulence, etc. because of the periodic boundary condition.

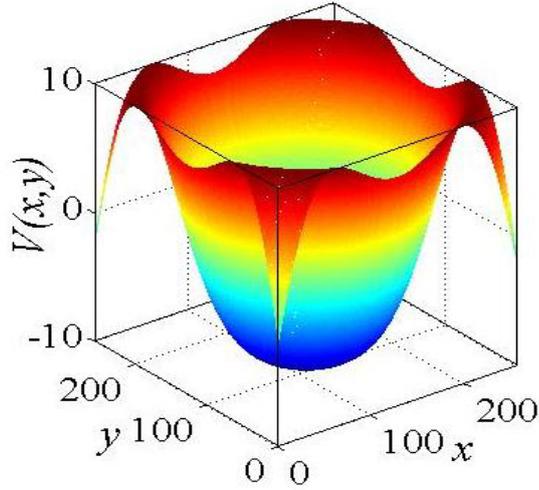

Fig.1. (Color online.) The external potential $V(x,y)$. Note that in this as well as the following figures the color is for clarity only. The values of the quantities are given by the vertical axis.

## A. Generation and spreading of turbulence

We first consider the case where the coefficients are real, say $p=1$ and $q=-1$, corresponding to a standard NSE with an external potential. The initial perturbation is of the form of a Gaussian pulse $E(0,x,y) = E_0 \exp\left[-\left(x^2+y^2\right)/R_0^2\right]$, where $E_0 = 30$ and $R_0 = 1.1$. That is, the large amplitude initial perturbation $E(0,x,y)$ is at the center of the simulation box and has a narrow spatial profile with respect to the latter.

In Fig. 2 we present the evolution of the energy spectrum $\left|E(k_x,k_y)\right|^2$ at $t$ = 0.005, 0.010, 0.020, 0.030, 0.050, 0.080, 0.100, and 0.200. One can clearly see that nonlinear modulation of the initial pulse first leads to a weak condensation in the $k$ space, followed immediately by side-band formation, which is in turn followed by a relatively slow cascading of the scale of the generated waves towards that of shorter and shorter wavelengths (larger $k$ values) until a fairly turbulent state appears. That is, the onset of turbulence is gradual and there is apparently no collapse, even though the initial perturbation is of very large amplitude.

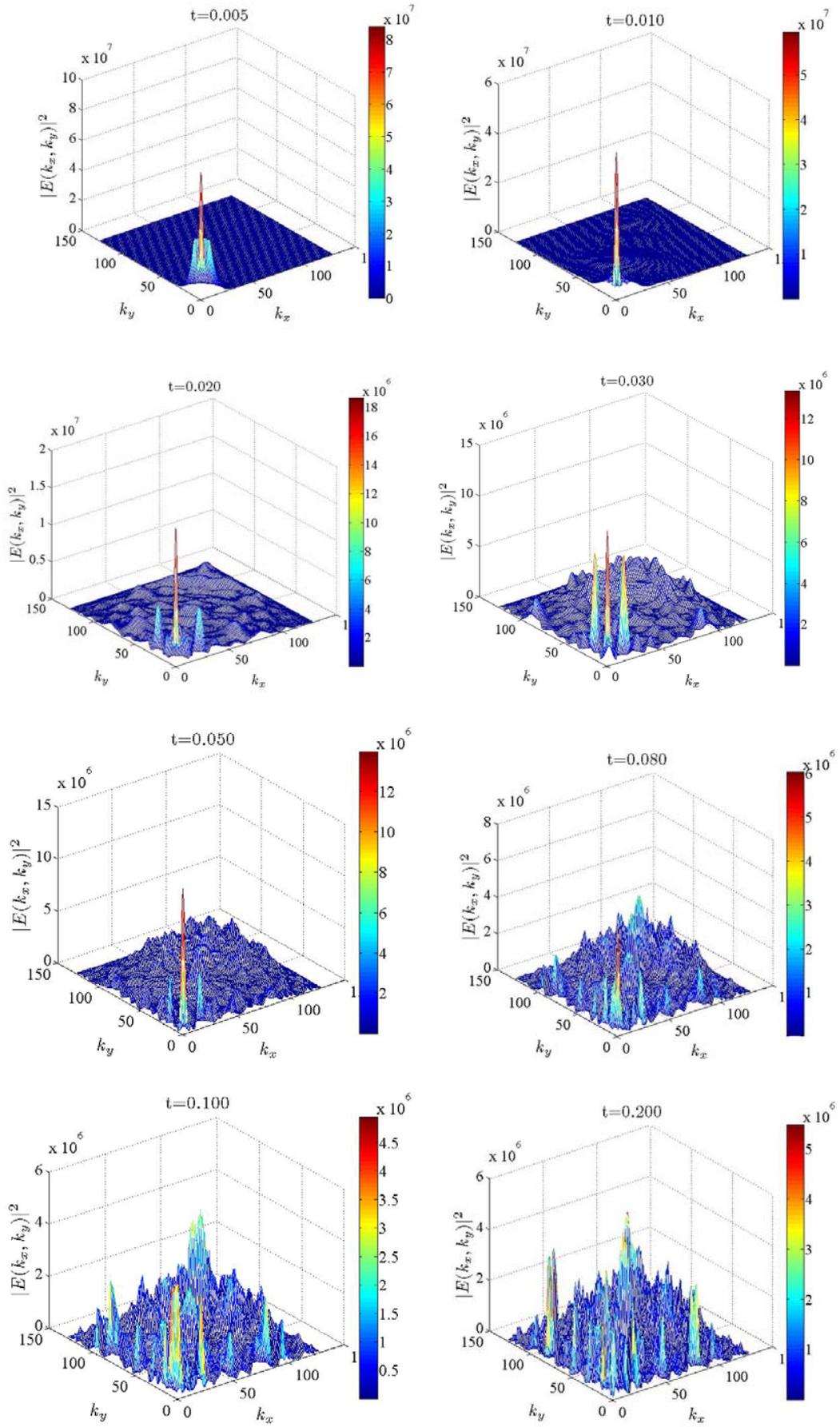

Fig. 2. (Color online.) Evolution of the energy spectrum of for the case $p=1$, $q=-1$, $R_0=1.1$, and $E_0=30$.

For completeness, in Fig. 3 we show the corresponding evolution of the enstropy spectrum $|kE(k_x,k_y)|^2$, where $k=\sqrt{k_x^2+k_y^2}$, at $t$ = 0.005, 0.010, 0.030, and 0.200. One can see that the enstrophy spectrum evolves similarly as the energy spectrum, but at different rates.

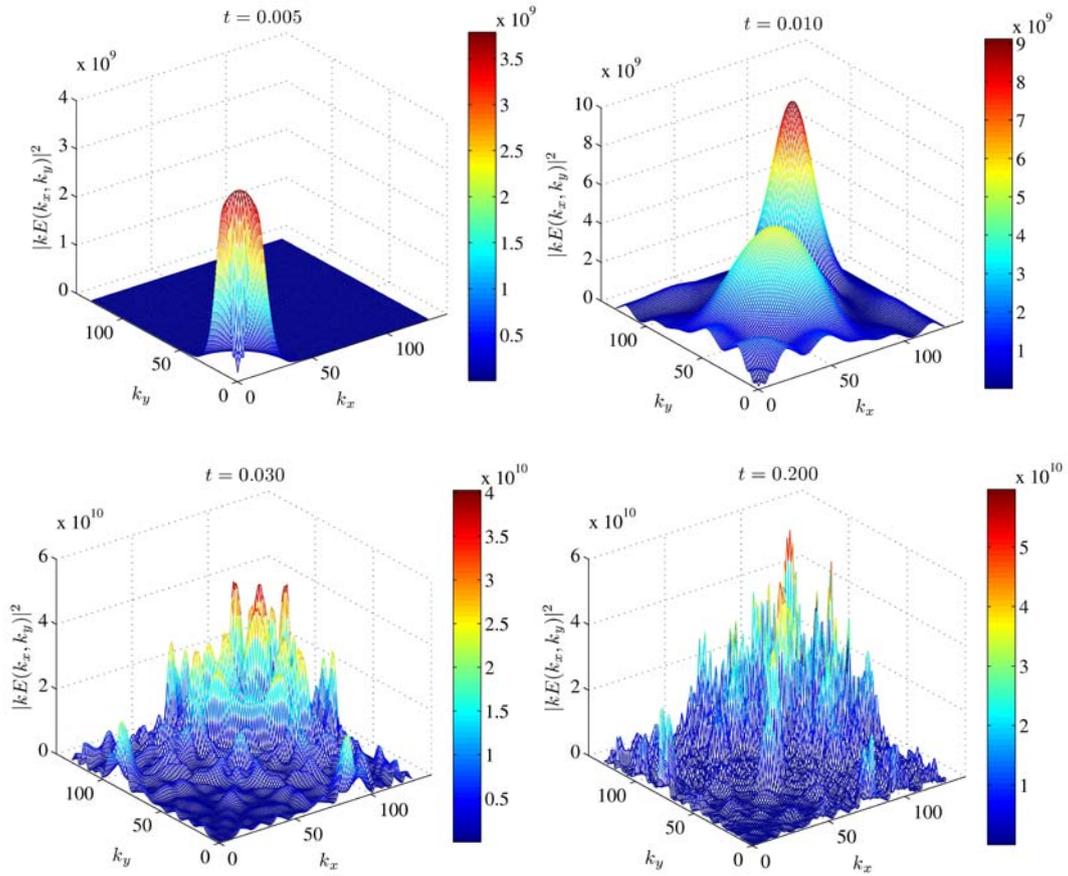

Fig. 3. (Color online.) Evolution of the reduced enstropy spectrum. The parameters are the same as in Fig. 2.

### B. Collapse and inverse cascade

Next we consider the case $p=-1+0.01i$, $q=10-0.5i$. That is, the system is characterized by negative group dispersion and viscous (or diffusive) growth, as well as amplitude-dependent damping. The initial Gaussian perturbation is of amplitude $E_0=1.6$, and of the same width $R_0 (=1.1)$ as the preceding case. The viscous growth coefficient is taken to be sufficiently small, and the nonlinear damping sufficiently large (recall that $E_0 = 1.6$), so that linear or nearly-linear growth of the initial perturbation is suppressed, which may overshadow the nonlinear evolution that we are interested in.

Figure 4 shows the energy spectrum corresponding to the evolution of the initial perturbation at $t$ = 0.150, 0.170, 0.200, 0.240, 0.255, 0.265, 0.275, 0.300, and 0.322. Here one can see that the initial perturbation also first condenses in the *k* space. However, instead of side-band formation and gradual spreading of the waves to smaller scales, at about $t$ = 0.2 the energy (actually a part of it, since the system is dissipative) in the long (small *k*) waves making up the initial pulse is abruptly (i.e., in a much shorter time scale) transferred to the very short (large *k*) waves, followed by gradual inverse cascade back toward the longer waves. Eventually a fairly uniform turbulent state is formed. That is, here we have a typical collapse scenario as predicted by Zakharov [1,4,5] who analytically considered the conservative case ($p_i$, $q_i = 0$) up to the singularity, and predicted that dissipation or other effects will then lead to short-wave turbulence. It should be noted that compared to the preceding case of a conservative system, here the initial perturbation is of relatively small amplitude. One can also see that the resulting turbulent state (which is not stationary) is much more spiky than that of the preceding case. This behavior can be attributed to the fact that here the turbulence is formed by harmonic interaction of the very short waves created at the end of the collapse stage (or, the start of the inverse cascade process), and in the preceding case it is formed by smooth modulation of the long waves of the initial pulse.

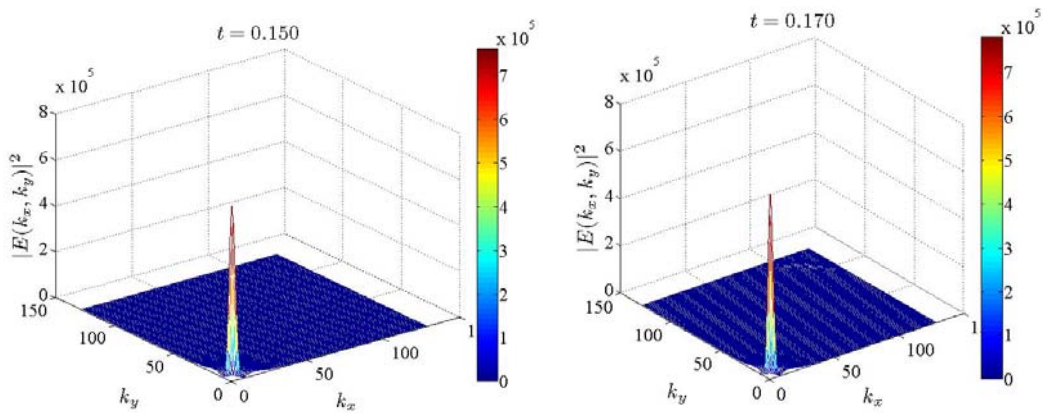

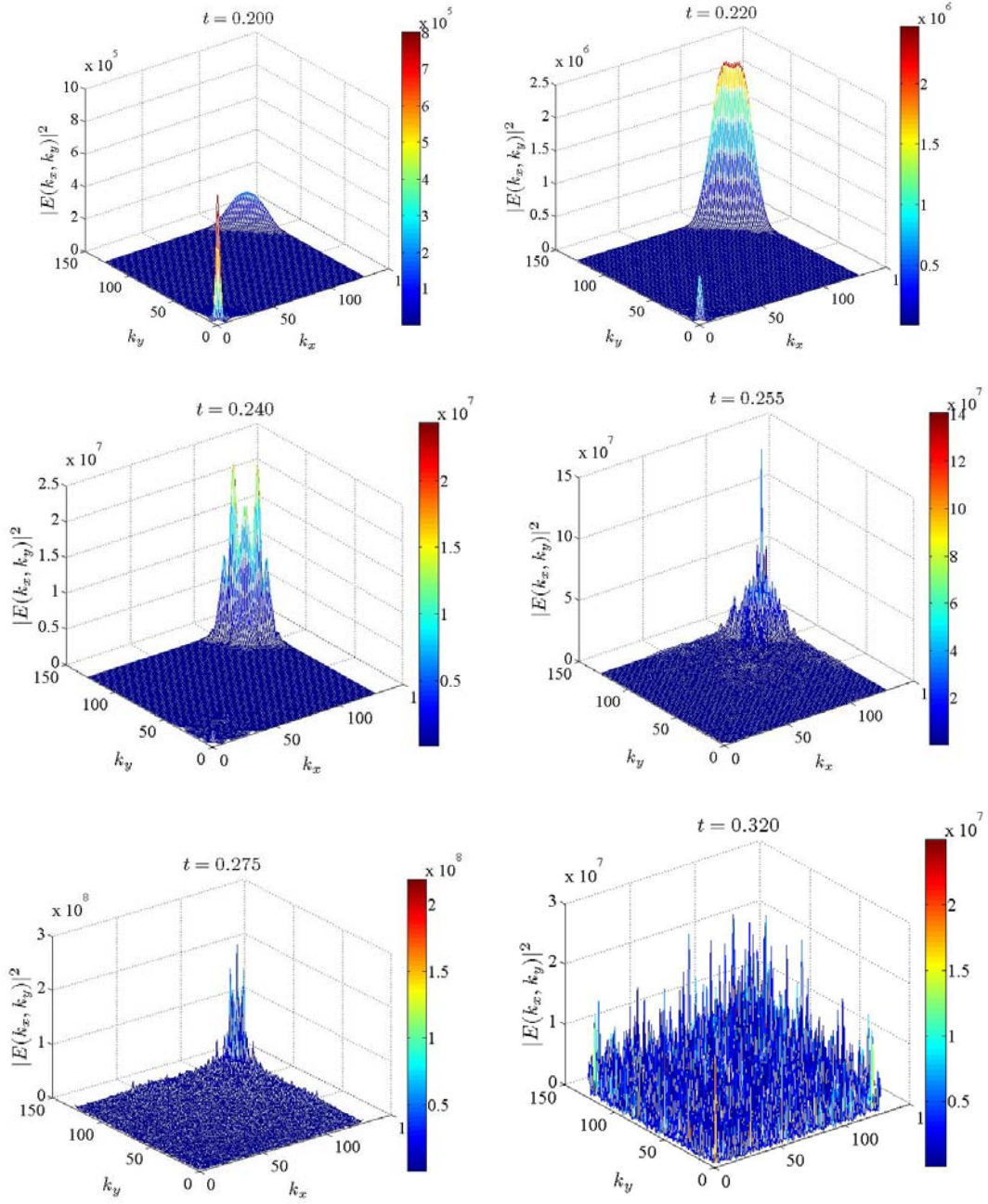

Fig. 4. (Color online.) Evolution of the energy spectrum of for $p = -1 + 0.01i$, $q = 10 - 0.5i$, $R_0 = 1.1$, and $E_0 = 1.6$.

Figure 5 shows the corresponding evolution of the enstropy spectrum $|kE(k_x, k_y)|^2$ at $t = 0.150$, 0.220, 0.255, and 0.320. One can see that again the enstropy spectrum evolves similarly as the energy spectrum, but at different rates.

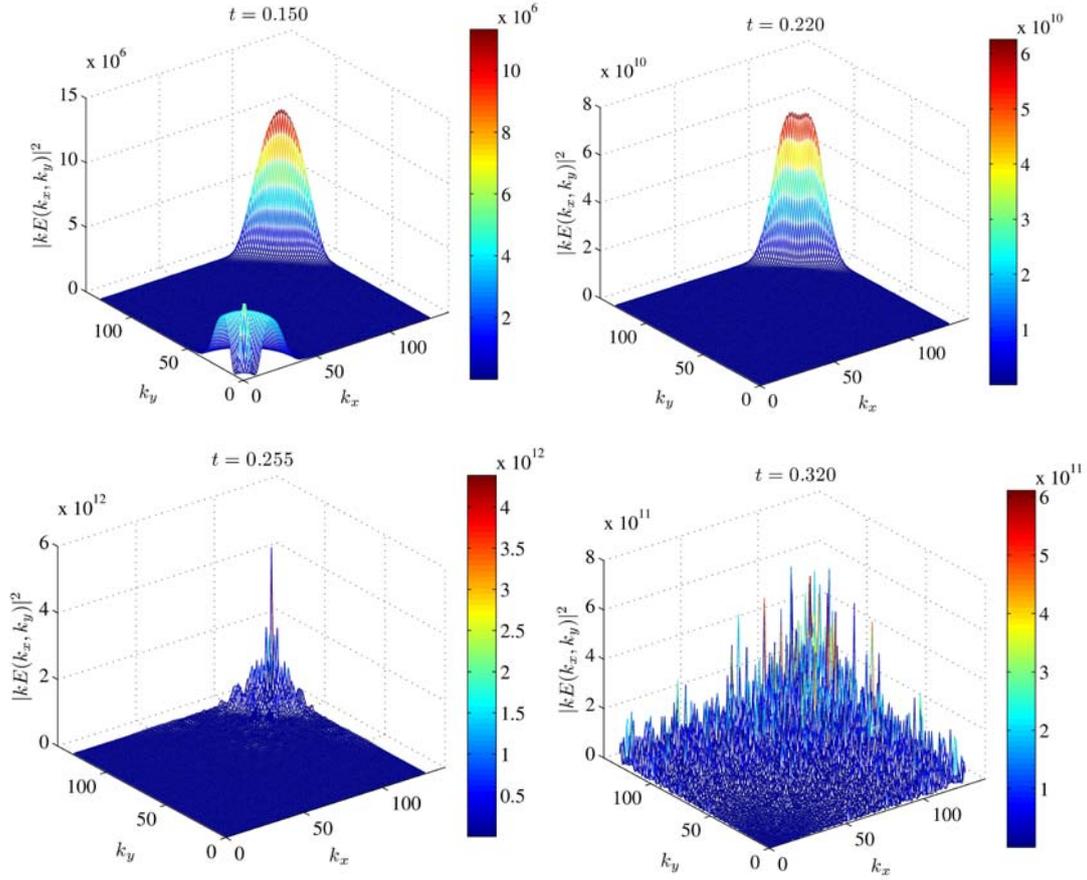

Fig. 5. (Color online.) Evolution of the reduced enstropy spectrum of for the same parameters as in Fig. 4.

## IV. Discussions

Although there is clear evidence of inverse cascade in most cases that we have considered, we have not found any evidence of self-organization, that is, the formation of regular or coherent structures from an earlier disordered state, despite the fact our system allows built-in viscous and nonlinear damping or growth, as well as management of energy and enstropy. We found that for the cases with nonlinear dissipation, the turbulence becomes more and more homogenous (but not stationary, as to be expected from Eqs. (3) and (4) for the total energy and enstropy) as the evolution continues. However, still no general conclusion can be made on this point since our numerical trials are necessarily limited in scope and not all of the many possibilities could be considered. More analytical investigation in this direction is clearly needed. On the other hand, with proper (such as introducing time and space dependent, or even feedback) control of the coefficients, one can use the generalized NLSE to investigate in great detail (e.g., at the different stages) the properties of wave cascade as well as possible control of turbulence.

Although the generalized NLSE can take into account features in nonlinear processes that are otherwise precluded, it should however be cautioned that for some physical systems the introduction of complex coefficients can lead to violation of the system's physical requirements, such as that of normalization and conservation. On the other hand, some physical systems can be investigated directly by numerical simulations based on first principles. For example, Langmuir wave collapse can also be investigated by solving the Vlasov equation numerically [22,23] and by particle-in-cell simulations [24]. It should thus be possible to verify the present results using these approaches after appropriate modifications. Finally, it may be of interest to point out that, because $p$ and $q$ are complex, the generalized NLSE can be expressed in other familiar forms, such as that not containing $i=\sqrt{-1}$ in the time derivative or the other coefficients. That is, the generalized NLSE is also relevant to completely different physical phenomena, such as that of the reaction-diffusion type that is common in the study of chemical reactions and biological evolutions [25–29], as well as in future theories on the development of economies [29] and financial markets [30].

**Acknowledgments**

This work was supported by the National Natural Science Foundation of China (10835003), the National Hi-Tech Inertial Confinement Fusion Committee of China, the National Basic Research Program of China (2008CB717806), and the Ministry of Science and Technology of China ITER Project (2009GB105005).